# Strain Analysis of a Chiral Smectic A Elastomer

Christopher M. Spillmann,[1,*] John H. Konnert,[1] James M. Adams,[2] Jeffrey R. Deschamps,[1] Jawad Naciri,[1] and Banahalli R. Ratna[1]

[1]Center for Bio/Molecular Science and Engineering, Naval Research Laboratory, Washington, DC, 20375, USA; [2]Physics Department, University of Surrey, Surrey, GU2 7XH, UK

* christopher.spillmann@nrl.navy.mil, phone: 202 767 0477, fax: 202 767 9594

**Abstract:**

We present a detailed analysis of the molecular packing of a strained liquid crystal elastomer composed of chiral mesogens in the smectic A phase. X-ray diffraction patterns of the elastomer collected over a range of orientations with respect to the X-ray beam were used to reconstruct the three-dimensional scattering intensity as a function of tensile strain. For the first time, we show that the smectic domain order is preserved in these strained elastomers. Changes in the intensity within a given scattering plane are due to reorientation, and not loss, of the molecular order in directions orthogonal to the applied strain. Incorporating the physical parameters of the elastomer, a nonlinear elastic model is presented to describe the rotation of the smectic-layered domains under strain, thus providing a fundamental analysis to the mechanical response of these unique materials.

**INTRODUCTION**

Liquid crystal elastomers are a unique class of material that couple the weak molecular ordering of liquid crystal phases to an underlying polymer network. Phase transitions and/or molecular reorientation from external fields have demonstrated that these materials can undergo reversible shape change [1-7]. Recent work has led to the development of chiral smectic A elastomers capable of macroscopic actuation via the electroclinic effect [5, 6, 8], a phenomenon wherein application of an electric field produces a molecular tilt in a plane orthogonal to a plane defined by the smectic layer normal and the transverse component of the permanent molecular dipole [9]. We have also recently provided a detailed analysis of the molecular packing and reorientation by observing the X-ray scattering intensity of the elastomeric material in the absence of external forces and also in the presence of applied electric fields or strain [10].

In the course of observing the many unique properties of smectic elastomers, one interesting aspect was the manner in which the material responds to an applied strain [3, 7, 11-16]. X-ray scattering experiments by Nishikawa and Finkelmann revealed a significant loss of the scattering intensity associated with the primary smectic layers of elastomer samples when subjected to tensile strain applied parallel to the director orientation [3]. It was suggested that the loss of intensity could be attributed to the smectic phase of the elastomer melting into a nematic-like state above a critical strain value. In a different smectic elastomer system subjected to strain perpendicular to the layer normal, Stannarius et al. also observed a loss of scattering intensity and speculated it was likely a decrease in the smectic layer order [13]. These reports and subsequent observations of similar systems were made in either a single plane or at most a few planes of orthogonal X-ray scattering intensity [3, 5, 7, 11-14], thus limiting the extent to which changes in the intensity could be attributed to the ordered domains potentially rotating out of favorable diffraction conditions. Adams and Warner later suggested that the energy cost of melting from a smectic to a lower-energy phase was rather large compared to rubbery elastic energy of the underlying network [17]. Based on a non-linear approach to model smectic A elastomers, a more likely explanation than a phase transition was rotation of the smectic layer normals out of favorable diffraction conditions and away from the direction of applied strain [17]. Subsequent work by Obraztsov et al. has analyzed the smectic line shape in different smectic A monodomains and revealed strong nonuniform strains and randomly distributed dislocations within the sample [18].

Only recently have we had a thorough look at multiple X-ray scattering planes of a smectic elastomer [10]. Reconstruction of the scattering intensity of an elastomer sample was obtained with no external strain or electric field applied, offering novel insight into the molecular packing arrangement in these thin films. The X-ray intensity of single scattering planes were collected with the sample films subjected to mechanical or electric fields which provided a tantalizing glimpse into the molecular reorientation. Knowing the limited amount of information that is collected from a single scattering plane, we now use a sample holder on an X-ray goniometer that simultaneously allows an elastomer film to be strained and rotated through multiple diffraction planes. Thus, the scattering information of strained elastomer films can be reconstructed to build a nearly complete three-dimensional view of the smectic layer-related scattering and offer unprecedented detail to changes in the smectic layer spacing and molecular orientation with respect to the direction of the applied strain.

In this paper we present the three-dimensional scattering intensity of a strained smectic elastomer and relate it to the order and distribution of the smectic layer domains. We find that the total integrated scattering intensity does not change as the sample is stretched and conclude there is no change in the smectic order but rather a redistribution of the domains with respect to the applied stress. The behavior of the strained elastomer is modeled using a nonlinear elastic approach first proposed by Adams and Warner [17] and later developed by Stenull et al [19].Taken together, we present a fundamental analysis and model of a smectic elastomer response to a mechanical field by observing the ordered domains in multiple dimensions.

## I. SAMPLES AND METHODS

### A. Elastomer preparation

The acrylate materials used to create the liquid crystal elastomer and a schematic of the polymerized network are shown in **Fig. 1(a)** and **(b)**, respectively. Synthesis of the materials and the preparation technique used to create free-standing films has been reported previously [6, 10]. In brief, a eutectic mixture of monoacrylate liquid crystal, 70 weight % of ACKN11 and 30 weight % ACBKN11, was combined with an additional 5 mole % of the diacrylate crosslinker DACP11 and a fractional amount of the photoinitiator Lucirin TPO. The mixture was filled into a 60 $\mu m$-thick glass template. Samples were aligned by slow cooling and exposure to an electric field of $\pm 6\,V/\mu m$ at 0.5 Hz. Once aligned in the smectic A phase, the field was removed and the sample photo-polymerized under UV light. The elastomer film was removed from the glass template by dissolving a sacrificial alignment layer in water. Samples were cut into strips ~ 20 mm in length and 6 mm in width.

### B. X-ray diffraction experiments

For collecting x-ray diffraction data on strained samples, an elastomer strip was mounted in a Huber goniometer head (Model #801.015). This head is used as a stretching device for films, foils, and wires. It consists of a horizontal clamp such that two ends of the film were secured. One end of the clamp was fixed while the other could be positioned to a desired distance to produce a measurable strain. The elastomer was mounted with the film initially flat and the clamp to clamp distance measured for the initial length, $l_0$. A set of diffraction data was collected at $l_0$ and subsequent data sets were collected as the sample was elongated in 0.25 mm increments. The data reported here is from a sample with $l_0 = 18.5$ mm and stretching the film by 0.25 mm introduced a 1.35% strain. Data sets were collected until the sample ruptured under 9.5% strain.

Diffraction data was collected using $CuK\alpha$ radiation and a Bruker Platinum-135 CCD detector on a MicroSTAR-H generator equipped with Helios optics. Samples were mounted in the goniometer clamp and initially positioned with the $y, z$ plane normal to the incoming X-ray beam, which was defined as $\phi = 0°$ (**Fig. 1(c)**). Prior to data collection, the goniometer was rotated about $\phi$ to determine the maximum angle the sample and holder could be moved without interfering with the collimating lens.

X-ray diffraction data sets were collected about $\phi$ in $1°$ increments from $-45°$ to $+45°$. This range was not sufficient to capture the entire scattering information of the elastomer sample, particularly the wide angle information related to the average intermolecular distance between adjacent molecules. However, it did allow for complete capture of the scattering information related to the primary smectic layer of the elastomer. See the Supplementary Information for movies of the diffraction data set collected in $1°$ increments about $\phi$ [20].

A full three dimensional reconstruction of the diffraction information was assembled using in-house software. Each 2D pattern was obtained as a $1024 \times 1024$ array of pixels for offline analysis. Background scattering of the images was subtracted using the method described in the Appendix. The three dimensional pattern for the elastomer under each strain was assembled and stored as a 512×512×512 array of voxels, allowing it to be sampled and displayed in any orientation. In addition, two dimensional slices were then extracted from the reconstruction to compare the scattering information along different planes relative to the axes of the elastomer defined in **Fig. 1(c)**.

## II. RESULTS

The majority of X-ray studies of LC elastomer films rely upon analysis from a single scattering plane, which is most often parallel to the surface of the elastomer film [3, 5, 15]. While this may be sufficient to understand a broad view of the molecular packing, specifically the layer-related information in smectic elastomers, it is only a small fraction of the elastomer scattering information. A more powerful approach to understand the molecular packing in liquid crystalline elastomers is to collect diffraction data over a range of sample orientations with respect to the X-ray source, thus enabling fundamental insight into the three dimensional organization of these materials.

As a first step, we present a 2D diffraction image of the elastomer collected in the $y, z$ plane parallel to the film surface (**Fig. 2(a)**). As we have shown previously [10], the scattering data can be categorized as two regions, the wide angle intermolecular spacing and the small angle smectic layer spacing. These two regions are orthogonal to one another, highlighting the presence of a smectic A phase in the elastomer. We focus our attention to the innermost scattering information associated with the primary smectic layer within the elastomer, which shows a distribution of intensity. Considering only the left half of the scattering in **Fig. 2(a)**, detailed analysis of the layer-related information reveals two predominant peaks deviating $\Box\, 20°$ from the $z$-axis and indicate the presence of two preferred chevron-like domains with an angular distribution centered $\pm 20°$ along the film width ($y$ axis). As shown in **Fig. 2(b)**, the same information is highlighted in a 2D slice of the $y, z$ plane taken from a reconstruction of the 3D scattering intensity collected in a $\phi$ scan of the unstressed sample. When a 2D slice is now taken in the $x, z$ plane from the reconstructed 3D scattering, the $20°$ deviation from the $z$ axis is still present, but there is a six-fold increase in the scattering intensity (**Fig. 2(c)**). These data emphasize the fact that the majority of the ordered smectic layer normals are tilted away from the axis of symmetry ($z$ axis) with most tilted toward the film normal ($x$ axis). The origin of the chevron domains in these films is likely a byproduct of the slow AC field alignment procedure of the monomeric mixture. This alternating chevron tilt or "sawtooth distortion" has been observed previously in smectic liquid crystals subjected to electric fields [21, 22], and in the present study this geometry is fixed by photopolymerization and cross-linking of the sample. We note that the black areas in **Fig. 2(c)** represent missing data from the limited rotation of the sample about $\phi$. The ability to slice the reconstructed data along different planes illustrates the power of capturing multiple frames of the elastomer at different orientations. It has allowed the Bragg condition for diffraction to be met in all orientations in order to construct a complete 3D view of the scattering intensity of the primary smectic layer of the liquid crystal elastomer.

The distribution of scattering intensity between the $x, z$ and $y, z$ planes is clearly shown when the 3D data is presented as a hemispherical cap instead of individual planes. If one considers rotating **Figs. 2(a)** and **2(b)** so that the $z$ axis extends through the center of the cap and out of the page over a spherical shell of $40 Å$ scattering intensity, the result is a pole plot as shown in **Fig. 3**. In polar coordinates, the shell includes a range of $r$ to include the scattering related to the $40 Å$ feature (**Fig. 3(a)**, inset) and the pole plots represent one pole of a hemispherical cap of the scattering intensity. In this reconstructed 3D image of the unstrained elastomer scattering intensity, the horizontal axis is the width of the sample ($y$ axis) and the vertical axis corresponds to the film thickness ($x$ axis). **Fig. 3(a)** represents a toroid-like distribution of the scattered intensity observed for the $40\text{Å}$ feature with the majority of scattering intensity distributed about the $x$ axis. The pole plot highlights the presence of the chevron-like domains, since the absence of these deviations would result in the scattering intensity tightly distributed on the $z$ axis at the center of the graph.

As the sample was stretched, the general shape of the scattering intensity remained the same but with a pronounced spreading away from the $z$ axis (**Fig. 3(b)**). A series of pole plots of the elastomer under increasing strain are provided in the Supplementary Material [20]. Relating the scattering intensity to the material response, the spreading of the $40\text{Å}$ scattering indicates a rotation of the chevron-like domains in a direction perpendicular to the applied strain and into the $x, y$ plane. This behavior was predicted by Adams and Warner [17] and here we provide the first direct evidence of smectic layer rotation in a direction perpendicular to the applied mechanical field. In polar coordinates, the rotation of the smectic layer domains, i.e. increase in the chevron angle, can be quantified as the change in $\phi$ as a function of strain and viewed as the deviation of the layer normal from the $z$ axis. **Fig. 4(a)** shows the average increase in the chevron angle as a function of strain. The angle was determined from the maximum intensity values of the $40\text{Å}$ scattering. The $\sim 15-20°$ increase in the chevron angle agrees well with our previous report [10]. It is important to highlight that if only the $y, z$ scattering plane had been observed, the increase in the chevron angle would be accompanied with a decrease in the scattering intensity.

The reconstructed scattering intensity of the $40 \text{Å}$ data allows detailed analysis of the d-spacing of the primary smectic layer as a function of strain, as shown in **Fig. 4(b)**. Taking the average d-spacing from the four intensity maxima (two from each Friedel pair), a relatively sharp decrease in the smectic layer spacing is observed as the elastomer is first strained by 1.35%, after which the d-spacing approaches $39 \text{Å}$ with additional strain. Overall, there is a ~ 2% decrease observed in the layer spacing as the sample is stretched. The reconstructed 3D scattering also allows the integrated intensity of the $40 \text{Å}$ scattering to be evaluated as a function of strain. This is an important parameter in determining if there are changes to the smectic order of the elastomer under applied strain. Maintenance of the integrated scattering intensity of the strained elastomer from the 3D data sets would provide the experimental evidence as to whether the ordered phase behavior was maintained and redistributed or being lost into a nematic or isotropic phase. As shown in **Fig. 4(c)**, the total intensity of the $40 \text{Å}$ scattering is preserved to a remarkable degree over the range of strains applied along the $z$ axis of the elastomer. We note that the next strain step at 9.5% is not included since the film ruptured during data collection. Taken together with the results in **Fig. 3(a)** and **3(b)**, we provide the first direct experimental evidence that the ordered smectic domains are being reoriented as the elastomer is strained.

The goniometer $\phi$ rotation over $90°$ allowed for capture and reconstruction of the entire $40 \text{Å}$ scattering information related to the primary smectic layer of the elastomer. The $40 \text{Å}$ feature occurs at small scattering angles, which equates to a large volume of scattering intensity being captured as the sample is rotated with respect to the X-ray source and detector. Put another way, if one considers the Ewald sphere with a radius of $1/\lambda$, where $\lambda$ is the wavelength of the incident radiation ($1.54 \text{Å}$) and the sphere's center is the origin of diffraction (elastomer) in real space, very small scattering angles can be approximated as falling nearly on a plane in reciprocal space. Since the detector records the intensity for a point in reciprocal space on the Ewald sphere, wider scattering features will have a greater amount of missing data given the limited elastomer orientation that could be sampled. This is true for the weaker second and third layer related features at $\sim 26\text{\AA}$ and $14\text{\AA}$. The data missing for the layer-related features are highlighted in the reconstructed scattering in **Fig. 2(c)**, which shows the convergence of the two Ewald sphere's from the $x, z$ plane. There were even greater limitations on the $4\text{\AA}$ information related to the average intermolecular spacing (see Appendix 2, Fig. A2), thus restricting the ability to reliably detect molecular tilt under applied strain or discern the presence of diverging populations of LC mesogens having two preferred tilt angles with respect to the long axis ($z$) of the elastomer. Despite these missing data, the reconstructed scattering intensity has allowed for an exceptional view of the primary smectic layer scattering information in order to better understand the molecular packing and response to external stimuli. Despite certain similarities observed in other smectic elastomer systems subjected to strain, notably a decrease in the smectic layer spacing often accompanied with a decrease in the primary smectic layer scattering intensity [3, 12, 13], it is possible that the strain response we have observed may differ from other systems due to significant differences in the composition and sample preparation. A model that incorporates the material parameters of our smectic elastomer is now presented to describe the experimental results observed under an applied strain.

## III. MODEL

We will use the model of smectic elastomers proposed in [17], and developed in [19], although phenomenological models have also been successful in modeling these materials [23]. A key assumption in these models is that the layers are convected by the rubber matrix, i.e. they deform like embedded planes, and there is no slip between the planes and the matrix. We denote the initial layer normal by $\mathbf{k}_0$, the rotated layer normal by $\mathbf{k}$ and the deformation gradient by $\lambda$. The layer normal is initially tilted at an angle of $20°$ with respect to the $z$ axis. Mathematically the assumption of embedded layers corresponds to

$$\mathbf{k} = \frac{\lambda^{-T} \cdot \mathbf{k}_0}{\left|\lambda^{-T} \cdot \mathbf{k}_0\right|} \qquad (1)$$

where the $-T$ superscript denoted the inverse transpose of the matrix.

The layer spacing $d$ between these embedded layers is given by

$$\frac{d}{d_0} = \frac{1}{\left|\lambda^{-T} \cdot \mathbf{k}_0\right|} \qquad (2)$$

where $d_0$ is the initial layer spacing.

We will consider one domain in the $x, z$ plane in the elastomer. The remaining domains are the same due to the uniaxial symmetry of the applied deformation. The experimental data does not support a diagonal deformation gradient, as the layer normal will not rotate fast enough with the imposed strain. We will assume that the deformation gradient has the form

$$\lambda = \begin{pmatrix} \lambda_{xx} & 0 & 0 \\ 0 & \dfrac{1}{\lambda_{xx}\lambda_{zz}} & 0 \\ \lambda_{zx} & 0 & \lambda_{zz} \end{pmatrix}, \qquad (3)$$

and that the domains in the sample have alternating shears that cancel out so that the macroscopic deformation is a pure elongation.

The free energy of the system will have a contribution from the nematic elasticity $F_{nem}$, the embedded smectic layers $F_{sm}$, and the coupling between the smectic layer normal and the director $F_{anc}$ given by the following expressions

$$F_{nem} = \frac{1}{2}\mu Tr[\lambda \cdot \ell_0 \cdot \lambda^T \cdot \ell^{-1}] \qquad (4)$$

$$F_{sm} = \frac{1}{2}B\left(\frac{d}{d_0} - \frac{\cos\theta}{\cos\theta_0}\right)^2 \qquad (5)$$

$$F_{anc} = \frac{1}{2}C(\cos^2\theta_0 - \cos^2\theta)^2. \qquad (6)$$

Here $\ell_0 = \delta + (r-1)\mathbf{n}_0\mathbf{n}_0$, and $\ell = \delta + (r-1)\mathbf{n}\mathbf{n}$ where $\mathbf{n}_0$ is the initial director, $\mathbf{n}$ the current director, and $r$ the polymer anisotropy. The rubber shear modulus is denoted by $\mu$, the smectic layer modulus by $B$, and the director-layer normal anchoring modulus by $C$. The tilt angle between the layer normal and the director is initially $\theta_0$, and $\theta$ in the current state, i.e. $\mathbf{n}\cdot\mathbf{k} = \cos\theta$. We have written the smectic layer spacing Eq. (5), in a slightly different form from [19] as it makes clear that $d$ is the layer spacing, and it is being driven by the rotation of the mesogens with respect to the layer normal. For physically reasonable values of $\theta$ close to $\theta_0$ there is no difference in the behavior of the model. A $\lambda_{zz}$ strain is imposed in the model and then the variables $\theta$, $\lambda_{xx}$, and $\lambda_{zx}$ are numerically minimized. The layer spacing and layer tilt can then be extracted as a function of imposed strain. Unfortunately the small deformation expansion of this model is algebraically complicated, and does not provide obvious guidance for data fitting. Typically the smectic layer modulus is at least a factor of ten greater than the rubber modulus $(B/\mu \sim 10)$. Here the elastic modulus of the sample is 4.0 MPa on stretching parallel $z$ axis, and 0.85 MPa on stretching in the perpendicular $z$ axis. Since the layers are slightly tilted with respect to the $z$ axis we can only deduce a lower bound on the value of the smectic layer modulus $B$ as we are stretching them at a small angle to their layer normal. However, there is experimental evidence that the difference between the rubber modulus and the smectic layer modulus is large, both from previous smectic samples [24] and from the high smectic modulus of $10^7 - 10^8$ Pa observed in liquid crystalline smectics. The model is not particularly sensitive to this parameter in the $B >> \mu$ regime. We will fit this model to the experimental data using the two parameters $C/\mu$ and $r$. The rate of rotation of the director with respect to the layer normal is governed by the polymer anisotropy $r$, and the modulus $C$. The larger $r$ and the smaller $C$ the faster director rotates with applied strain. Typically $C/\mu \sim 1$ and $r \sim 2$ for side chain liquid crystalline polymers [7, 19]. The defect structures in the smectic phase indicate that it is in the smectic A phase. To be consistent with this observation we will assume in our modeling that the tilt between the director and the layer normal in the initial state is $\theta_0 = 0$. Both the chevron angle and the layer spacing can be fitted simultaneously which imposes significant constraints on the model. Using the parameters $C/\mu = 0.2$, $r = 1.5$, and the experimentally measured quantities of initial tilt of the layer normal, and initial layer spacing, a rough fit to both plots can be obtained **Fig. 5**.

There is some compromise between the two plots to obtain these fits, for example a better fit to the chevron angle data can be obtained for a smaller value of the anisotropy $r$, however this produces a worse fit to the layer spacing. There is a noticeable bump at the start of **Fig. 5(a)** which results from our assumption that the sample is initially in the smectic-A phase such that $\theta_0 = 0$. As a result, the change in layer spacing, which is caused by the rotation of the director away from the layer normal, changes quadratically with strain at small strain. The sharp drop in the layer spacing for small strains could also be replicated with a non-zero $\theta_0$ value. The layer spacing would then be first order in the strain rather than second order. The subsequent data fit is almost identical. While it is reasonable to have a non-zero value of $\theta_0$ for de Vries-like materials, there are no data points in the 0-1% region to distinguish between these two cases. Consequently, only the $\theta_0$ plot is shown as it is compatible with the defect structures in the sample. The trend of the model chevron angle in **Fig. 5(b)** systematically overestimates the rotation angle. The faster rotation of the chevron angle with strain in the model could be explained by the lack of any semi-soft terms arising from a distribution of chain lengths in the rubber for example. In smectic elastomers these could take a variety of forms due to the different physical directions in the problem [25]. The assumption of the structure of the deformation matrix and that each domain can deform independently could also result in a better fit. These assumptions could be explored through a finite element model. Given that only two fitting parameters were used here, both with a limited range of values due to their physical interpretation in the microscopic model, the agreement between the data and the model is good.

**Figure 6(a)** shows the angle the director makes with respect to the stretch axis. Initially it is decreased as the director rotates toward the stretch axis, but later starts to rotate away due to the energetic penalty for deviating from the layer normal. Consequently it remains roughly constant. This is reasonable for the elastomer sample, but cannot be reliably corroborated with experimental evidence given the limited amount of scattering collected at $4\,\text{Å}$. The tilt angle $\theta$ between the director and layer normal is shown in **Fig. 6(a)** and initially has a sharp increase, which causes a corresponding layer spacing decrease. Subsequently, $\theta$ increases more gradually as the director starts to rotate around together with the layer normal. The rate of reduction in the layer spacing with strain then slows down. Physically it is energetically favorable for the director to be aligned with the strain direction, yet the direction of the layer normal is constrained to rotate as an embedded plane. Consequently, the tilt angle increases and the layer spacing contracts. However, the increased tilt angle is penalized by the $F_{anc}$ term in the free energy which eventually causes the director to start its rotation away from the strain axis. **Figure 6(b)** shows the $zx$ component of the shear, and the deformation in the $x$ direction as the sample is deformed. These are illustrated in **Fig. 6(c)** for a single domain in the film. The film contains many such domains distributed with cylindrical symmetry around the stretch axis.

**CONCLUSION**

We have collected and analyzed the primary smectic layer X-ray scattering intensity of a smectic A elastomer as as function of applied strain. Reconstruction of the three-dimensional scattering provides the first direct evidence of preservation of the smectic order and rotation of smectic domains in directions perpendicular to strain applied along the long axis of the sample. The experimental results are consistent with the assumption of smectic layers embedded in an affinely deforming rubber matrix. The layer normal of these embedded planes rotates away from the direction of the applied strain. The rotation of the director away from the layer normal causes the layer spacing to reduce with strain. Similar experimental techniques would prove useful in understanding the rotation of the layer normal in polydomain Sm-*C* samples [16]. Here the layer normals of the domains are distributed around a cone tilted at an angle $\theta$ with respect to the director. Future experiments will attempt to collect a more complete picture of the three-dimensional X-ray scattering intensity in order to reveal the reorientation of the layer normals as the sample is deformed.

**ACKNOWLEDGMENTS**

We thank the Office of Naval Research for funding support.

**APPENDIX 1: Background Calculation and Subtraction from the 3D Scattering Intensity**

The 3D diffraction data are stored in Cartesian coordinates $(x, y, z)$ that are linear in reciprocal space. The spherical coordinates of $(x, y, z)$ of the 3D intensity are given by

$$r = \sqrt{x^2 + y^2 + z^2} \qquad (A1)$$

$$\theta = \tan^{-1}(y/x) \qquad (A2)$$

$$\varphi = \cos^{-1}(z/r) \qquad (A3)$$

with the reverse transforms being

$$x = r \cdot \sin\varphi \cdot \cos\theta \qquad (A4)$$

$$y = r \cdot \sin\varphi \cdot \sin\theta \qquad (A5)$$

$$z = r \cdot \cos\varphi. \qquad (A6)$$

The spherical shells of data are computed and displayed in the coordinate system $(r, \theta_x, \theta_y)$ for which the spherical coordinates are given by

$$r = r \qquad (A7)$$

$$\theta = \tan^{-1}\left(\theta_y / \theta_x\right) \qquad (A8)$$

$$\varphi = \sqrt{\theta_x^2 + \theta_y^2} \qquad (A9)$$

We see that both coordinate systems $(r, \theta_x, \theta_y)$ and $(r, \theta, \varphi)$ define $r$ identically. All $r$ values of constant $\theta_x$ and $\theta_y$ fall on the same vector directed from the origin in reciprocal space through the $(x, y, z)$ point at which we evaluate the background value.

When analyzing the relatively sharp features of the elastomer 3D diffraction data, it is useful to subtract background contributions before evaluating positions of maxima, angular distributions and integrated intensities of these features. We describe the method used to subtract the significant background intensity from the scattering intensity related to the $40\text{Å}$ feature of the elastomer in a consistent manner for each set of scattering data collected at a given strain value.

**Determination of background end points**

The entire 3D diffraction pattern is used to determine the radii of the end points for the computation of the backgrounds for the individual points in the diffraction pattern. The radii values marked $r_1$ and $r_2$ (see **Fig. 7a**) are chosen as end points for the linear background calculation to be computed for each individual radial vector, $r_0$, of constant $\theta_x$ and $\theta_y$. These end points have coordinates $(r_1, \theta_x, \theta_y)$ and $(r_2, \theta_x, \theta_y)$. The variables $r$, $\theta$, and $\varphi$ are computed using Eqs. (A7) through (A9). These values are then used with Eqs. (A4) through (A6) to determine the data point $I(x_0, y_0, z_0)$ in the $512 \times 512 \times 512$ array of intensities that is closest to the location $(r_0, \theta_x, \theta_y)$. We wish to evaluate coordinates of the background end points that have the same $\theta_x$ and $\theta_y$ values as data point $(x_0, y_0, z_0)$ but with radii $r_1$ and $r_2$. These points lie along a vector from the origin of reciprocal space to point $(x_0, y_0, z_0)$. To compute the desired coordinates, we need only compute the ratios $r_1/r_0$ and $r_2/r_0$ with which to scale the coordinates $(x_0, y_0, z_0)$.

$$x_{11} = \frac{r_1}{r_0} x_0 \tag{A10}$$

$$x_{22} = \frac{r_2}{r_0} x_0 \tag{A11}$$

The variables $y_{11}$, $y_{22}$, $z_{11}$ and $z_{22}$ are defined in a similar manner. We find the data point $I(x_1, y_1, z_1)$ in the $512 \times 512 \times 512$ array of intensities that is located closest to the location $(x_{11}, y_{11}, z_{11})$. The indices of the intensity array are integers, with the nearest integer to $x_{11}$ being $x_1$. $I(x_2, y_2, z_2)$ is determined in the same manner.

The background intensity value, $I_b$, at a point $x_0$ is then computed as follows

$$slope = \frac{I(x_2, y_2, z_2) - I(x_1, y_1, z_1)}{r_2 - r_1} \tag{A12}$$

$$I_b = slope \cdot (r_0 - r_1) + I(x_1, y_1, z_1) \tag{A13}$$

The value $I_b$ is then subtracted from $I(x_0, y_0, z_0)$ to yield the intensity contribution, $I$, from the feature of interest

$$I = I(x_0, y_0, z_0) - I_b. \tag{A14}$$

When the projected intensity of a shell of data is then summed for intensity evaluation and plotting, each element for a given radial vector is multiplied by the squared radius of that element. Following subtraction of the background scattering, the data were evaluated to obtain d-spacings at various orientations (**Fig. 7b**).

**Computation of integrated intensity within a shell of reciprocal space**

The intensity within shells was also computed by carrying out the integration in Cartesian coordinate space. Since all elements in the 3D array represent equal volumes of reciprocal space, the integrated intensity of a feature is obtained by considering all elements within the chosen shell and summing all of their intensity contributions computed as described in the previous section.

**APPENDIX 2: Excluded wide angle scattering**

The geometry of the goniometer that held the strained elastomer sample limited the extent to which the sample could be rotated about $\phi$. As such, there is a significant excluded cone of scattering information. These missing data are shown in the 2D slice of the reconstructed 3D scattering through the *y, z* plane in **Fig. 8**. Since the majority of the $4\text{Å}$ scattering was missing, there are limited conclusions we can draw from these data. This is particularly true in regard to the presence of preferred domains of the director.

**Figures**

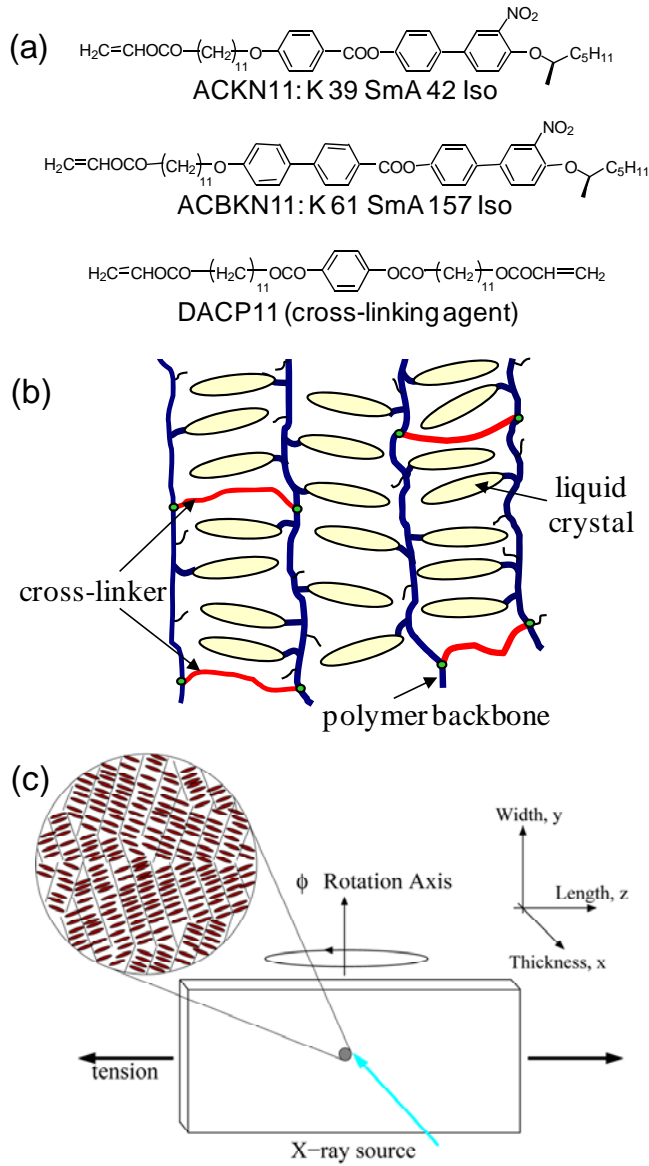

**FIG 1.** (Color online) Materials and elastomer setup for X-ray scattering. (a) Monoacrylate liquid crystal mesogens and diacrylate cross-linking agent. Phase behavior of monomeric materials is provided. (b) Schematic representation of elastomer network with pendent liquid crystal mesogens. (c) Orientation and rotation axis of elastomer with respect to the X-ray source. X-ray CCD detector position was fixed normal to the source. Chevron-like domains in elastomer are tilted with respect to the tension axis at a fixed angle in all directions.

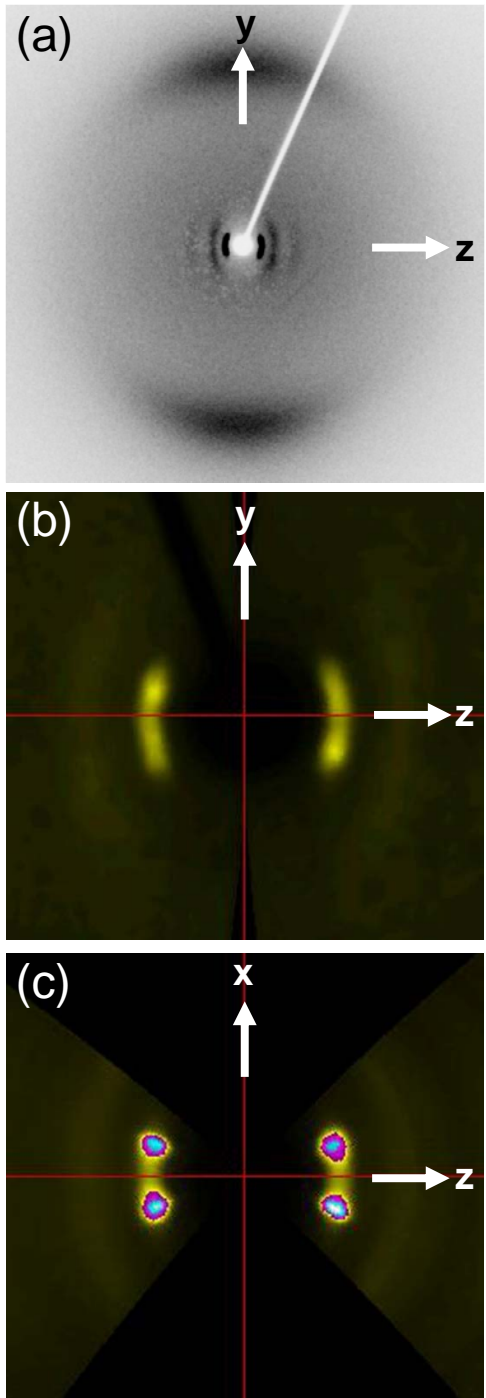

**FIG 2.** (Color online) X-ray scattering intensity of smectic elastomer. (a) 2D scattering collected in the $y, z$ plane showing the wide angle feature and the three small angle features related to the smectic layers. (b) Slice in the $y, z$ plane of the reconstructed 3D scattering showing the primary smectic layer. (c) Slice in the $x, z$ plane of the reconstructed scattering highlighting that the most intense scattering is located along this plane. Images in (b) and (c) have been scaled by the maximum intensity in (c).

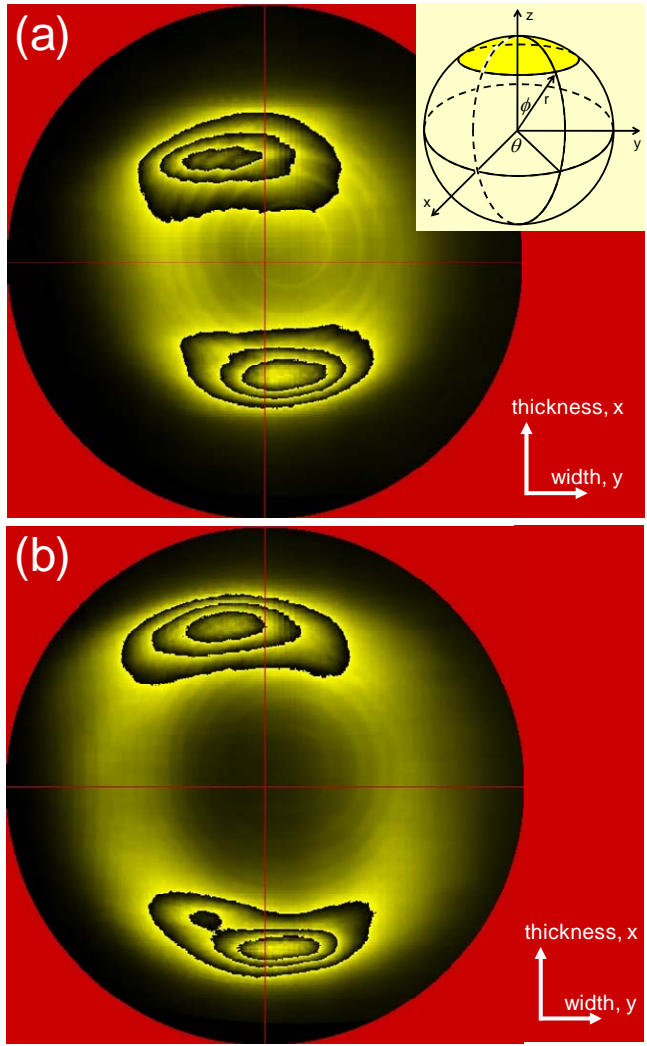

**FIG 3.** (Color online) Pole plots of the reconstructed scattering intensity of the $40\text{Å}$ feature under (a) 0% strain and (b) 8.1% strain. Spreading of the scattering away from the central point show the ordered domains rotating away from the strain applied along $z$. More intense scattering is shown in the upper and lower regions of the plots. Inset: upper yellow cap represents the portion of a sphere viewed along the $z$ axis in (a) and (b).

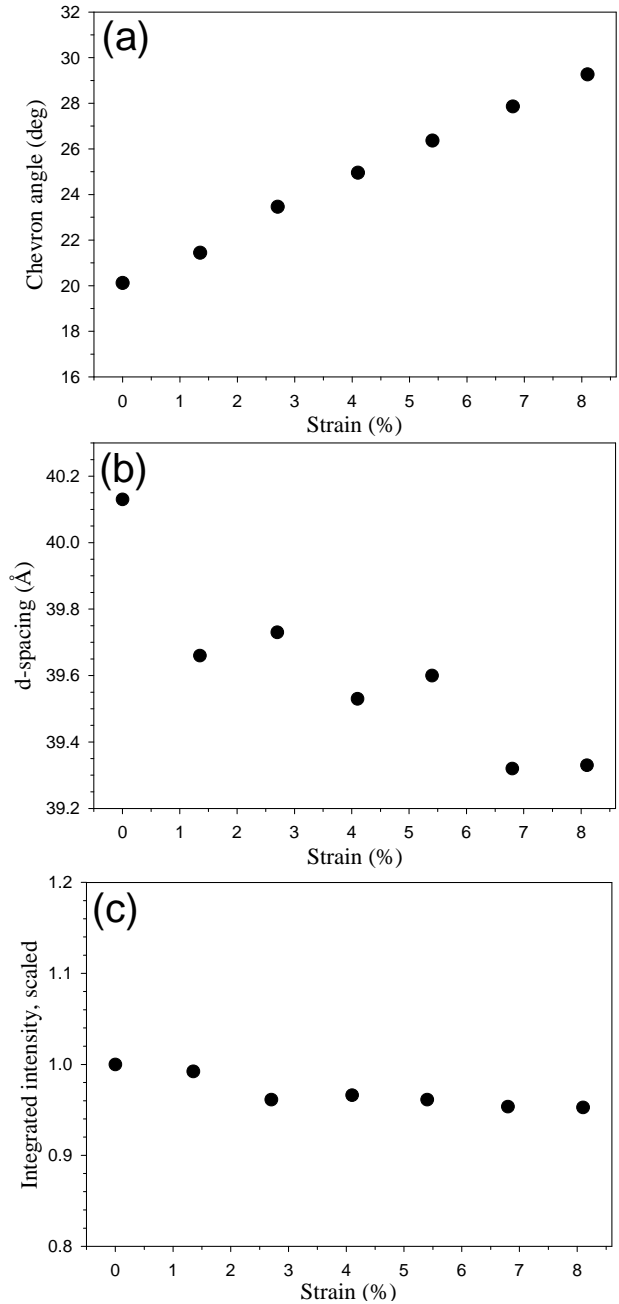

**FIG 4.** Analysis of $40 Å$ feature. (a) Chevron splitting, defined as the angular deviation from the $x$, $y$ plane, as a function of strain. (b) Change in the primary layer spacing as a function of strain. Data points were determined as the average of the maximum scattering intensity of the Friedel pairs and represent a 2% change in the d-spacing over the range of applied strains. (c) Integrated scattering intensity of the $40 Å$ feature showing preservation of the intensity as a function of applied strain. Data is scaled by the intensity measured under $0\%$ strain and includes the entire scattering information integrated about the spherical coordinate $\theta$ in Fig. 3(a), inset.

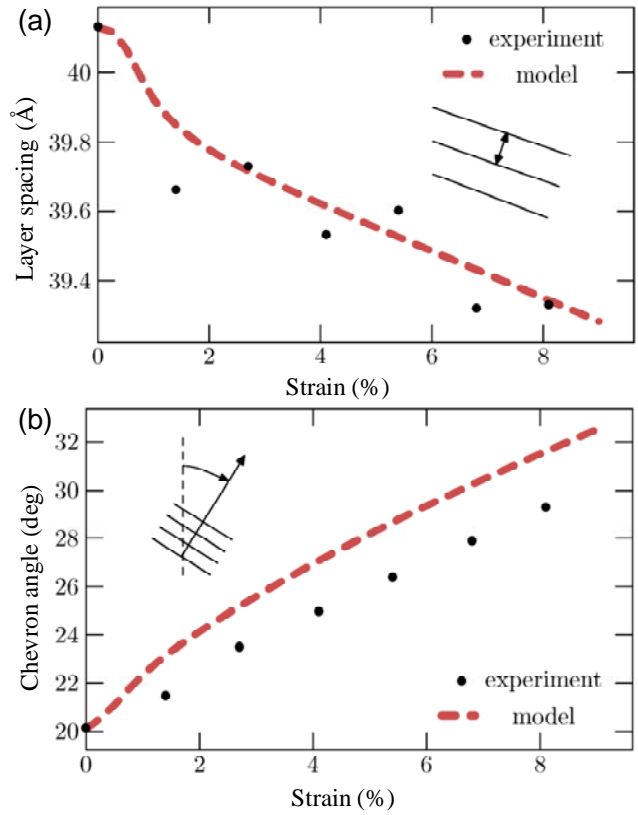

**FIG. 5.** (Color online) (a) Fit of the model to the layer spacing data, and (b) the chevron rotation data. Model parameters are provided in the main text.

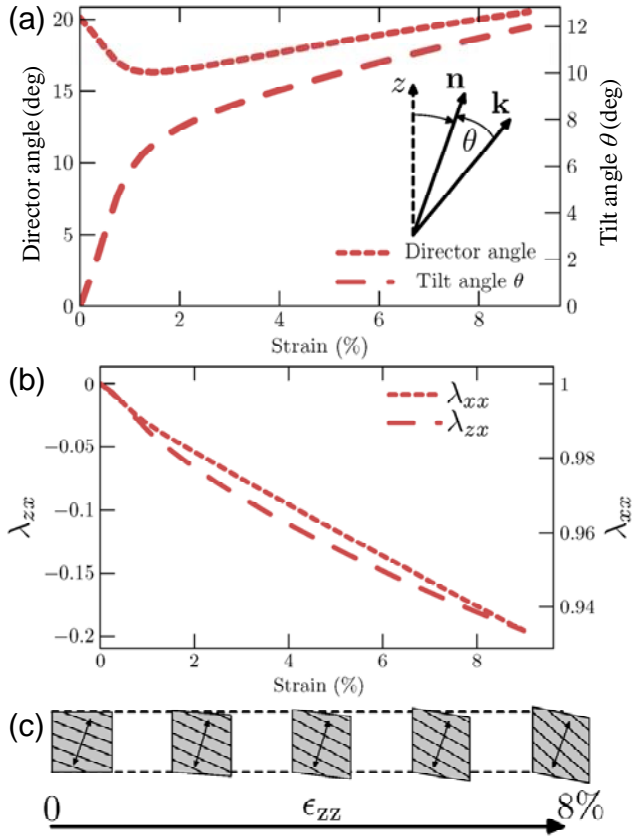

**FIG. 6.** (Color online) (a) The director angle with respect to the stretch axis, and the tilt angle as a function of applied strain. (b) The components of the deformation $\lambda_{xx}$ and $\lambda_{zx}$. (c) A scale cartoon of the layer rotation angle, the director and the layer rotation angle as a function of applied strain for a single domain.

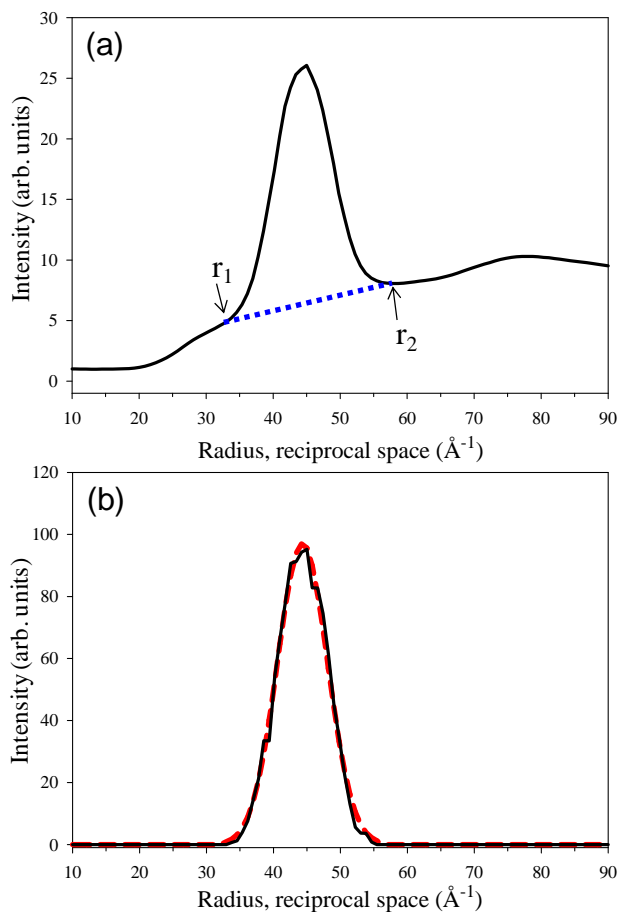

**FIG. 7.** (Color online) Background subtraction of the scattering intensity. (a) Radial average of the 3D intensity used to determine $r_1$ and $r_2$ in order to compute background values for individual radial vectors. (b) Gaussian fit (red dashed line) to experimental scattering intensity with background subtracted (solid black line). Data was used to obtain individual d-spacing for individual radial vector passing through the maximum.

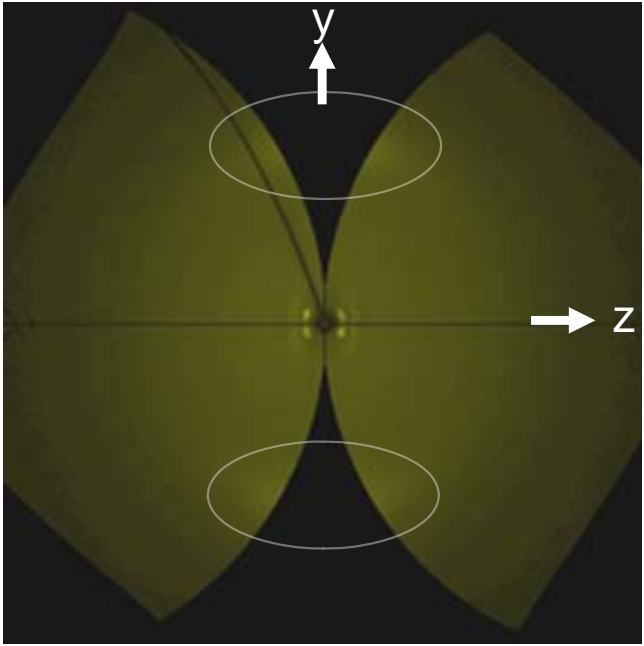

**FIG. 8.** (Color online) Cross-section of two Ewald's spheres showing the paucity of the $4\text{Å}$ scattering information collected in each data set. The black regions represent missing data and the open white ovals highlight the region of the $4\text{Å}$ scattering.